\documentclass[a4paper,11pt]{article}
\pdfoutput=1 

\usepackage{jheppub} 

\usepackage[T1]{fontenc} 
\usepackage{braket}
\usepackage{graphicx}
\usepackage{amsmath, amssymb}
\usepackage{bbold}
\usepackage{comment}
\excludecomment{confidential}
\usepackage{makecell}
\usepackage{bm}
\usepackage{mathtools}

\newcommand\args{(t,x,x')}
\newcommand\argsO{(0,x,x')}
\newcommand{\phicl}[2]{\phi_{cl}^{ #1}{( #2)}}
\newcommand{\cK}{ {\mathcal K}}

\newcommand{\AMSeqn}[1]{\begin{align} #1\end{align}}
\newcommand{\eqn}[1]{\begin{eqnarray}#1\end{eqnarray}}

\newcommand{\nn}{\nonumber}
\newcommand{\half}{\frac{1}{2}}
\newcommand{\threehalves}{\frac{3}{2}}

\newcommand{\quarter}{\frac{1}{4}}

\newcommand{\expect}[1]{\langle{ #1}\rangle}

\title{The Size of a Soliton}

\author[a]{John F. Wheater}
\author[a]{P. D. Xavier}

\affiliation[a]{Rudolf Peierls Centre for Theoretical Physics, Clarendon Laboratory, Parks Road,
University of Oxford, Oxford OX1 3PU, UK}

\emailAdd{john.wheater@physics.ox.ac.uk}
\emailAdd{praveen.xavier@physics.ox.ac.uk}

\abstract{We consider the recently proposed Bound State Conjecture for quantum field theory in the context of solitons in two dimensional $\phi^4$ scalar field theory. We calculate the quantum correction to the size of the soliton which, taken together with the known mass correction, shows behaviour consistent  with regarding the soliton as a bound state, and with the conjecture.
}

\keywords{soliton size, form-factor}

\begin{document} 
\maketitle
\flushbottom

\section{Introduction \label{bcconj}}

The requirement \cite{Arkani-Hamed:2006emk} of a consistent ultra-violet completion  including gravity is a powerful route to constraining field theories for low energy phenomena and has led to many significant results -- see \cite{Palti:2019pca} for a recent review and \cite{vanbeest:2021abc} for recent lecture notes on the topic. In particular the Weak Gravity Conjecture leads to constraints on scalar masses and coupling constants \cite{Palti:2017elp}.  These results motivated the recent suggestion
that the minimal size for a bound state system exists \emph{independently} of an ultraviolet completion including gravity.  The Bound State Conjecture  \cite{Hebecker2019} proposes  that bound states in a renormalizable quantum (effective) field theory (QFT) must have a minimum radius in the following sense.  
Consider a bound state of radius $R$ in a theory where $m\ll \Lambda$ is the mass of heaviest elementary particle and $ \Lambda$ is the cut-off scale; then the dimensionless quantity $mR=\Delta(\{\lambda\})$  is a function of the parameters of the theory, $\{\lambda\}$, such as the couplings and other particle masses.    
The proposal of \cite{Hebecker2019} is that $\Delta(\{\lambda\})$ is bounded below by some positive constant, i.e. $\Delta(\{\lambda\})\geq\delta>0$.
The conjecture is non-trivial because, in a theory with attractive interactions
one can imagine turning-up the appropriate coupling  indefinitely and squeezing any bound state to an arbitrarily small size. Clearly for systems that are only weakly bound, for example the Hydrogen atom, the size will initially decrease as the coupling strength is raised.
However, as this happens, heuristically the constituents become more energetic (momentum $p\sim R^{-1}$) and the process may become unstable 
when there is enough kinetic energy for particle production (similar to the mechanism involved in the Hagedorn limit \cite{Hagedorn}).

The authors of \cite{Hebecker2019} considered a number of examples, mostly in scalar field theories in four dimensions, where they were able to show  mechanisms that lead to the conjecture being satisfied. Field theories containing non-elementary states where the full dynamics can be closely controlled are not so common, but theories containing topological solitons are one example. 
Such solitons are known to possess a particle-like interpretation in QFT \cite{GervaisExtended1975}; on the other hand they are also extended objects for which stability against finite energy perturbations is guaranteed so $R$ should remain a well-defined quantity independent of coupling.  Solitons can also be viewed as a bound state of the fundamental point-like particles; for example baryons in $N_f>1$ QCD are described by the Skyrme model \cite{Zahed:1986qz}, \cite{ma:2016abc}, and in $N_f=1$ QCD by a quantum hall droplet \cite{Komargodski:2018abc}. 
In this paper we investigate the behaviour of the radius of kink solution of two-dimensional $\phi^4$ field theory (hereafter the $\phi_2^4$ model) by computing its first quantum correction as a first step toward understanding whether the conjecture applies for this model.

Let us briefly summarise the arguments and contents of this paper. Solitons are finite energy, non-dissipative solutions to the classical field equations --
non-dissipative means that
$\lim_{t\to\infty}\max_{\bm x}\mathcal{H}(t,\bm x)\neq 0$ \cite{Coleman1985} (Sec. 6.1), where $\mathcal{H}$ is the energy density.
Intuitively, this means that the field stays lumped-up for all time. Classically, the radius of a soliton, $R_{cl}$, is easy to define and compute: we adopt the definition that it is the width of the energy density, normalized by the total energy \cite{Lee1987},
\begin{equation}
    R^2_{cl}=\int d^d x \,\,|\bm x|^2 \mathcal{H}(\bm x)/\int d^d x\, \mathcal{H}(\bm x). \label{firsteq}
\end{equation} 
If the soliton solution is known then \eqref{firsteq} can be evaluated straightforwardly. To derive the quantum correction  requires a quantum theory of solitons and a method to compute matrix elements involving a single soliton in the initial and final states. The collective coordinate quantization developed  in the 1970's \cite{GervaisPoint1976,Raj} enables this.

The configuration space of QFT in $\mathbb{R}^{d+1}$ is the set of field configurations on $\mathbb{R}^d$. (Let $\bm x$ denote positions in $\mathbb{R}^d$.) An elementary particle state of the QFT is a plane-wave impulse of the field. This can be made precise: take the initial data $\phi_{cl}(\bm x)=\varepsilon\, e^{i\bm p.\bm x}$, $\pi_{cl}(\bm x)=\varepsilon\, ip^0e^{i\bm p.\bm x}$ and define an operator $O=\exp i\int d^d\bm x \left(\pi\phi_{cl}-\phi\pi_{cl}\right)$. This operator has the property that $O^{-1}\phi O=\phi+\phi_{cl}$ and $O^{-1}\pi O=\pi+\pi_{cl}$. Therefore, it causes a shift by the classical configurations $\phi_{cl}$, $\pi_{cl}$. In the limit that $\varepsilon$ becomes infinitesimal, $\partial_{\varepsilon}O|_{\varepsilon=0}=a^\dagger_{\bm p}$. Therefore, $a^\dagger_{\bm p}$ creates a plane wave impulse. In just the same way, if the QFT contains a soliton, described by some initial data $\phi_{cl}$ and $\pi_{cl}$, then we can construct a soliton creation operator (essentially a coherent state operator)
-- as was first done in \cite{Cahill1974}. 

To obtain a state that \emph{approximately} describes an elementary particle, one usually applies $a^\dagger$ to the vacuum state $\vert 0\rangle$. Similarly, to obtain a state approximately describing a soliton, one should act with $O$ on the vacuum. In an interacting theory these approximate states (hereafter called `proxy states') are not eigenstates of the Hamiltonian (which we will call `exact states'). However, the proxy states  are designed to have non-zero overlap with the corresponding exact state with the same quantum numbers. Assuming that we are dealing with the lowest energy state in the channel, the exact state can be extracted by a limiting procedure: $\ket{\text{exact}}=\lim_{T\to\infty}e^{ET}e^{-HT}\ket{\text{proxy}}/\braket{\text{exact}|\text{proxy}}$ where $E$ is the energy of the state. 
With this procedure in hand, one can study matrix elements between exact states. For example, given an exact elementary particle state $\ket{p^\mu}$ consider $\bra{p'}\mathcal{H}\ket{p}$.
Using the limiting procedure, this can be written as a matrix element between proxy states: $\sim\bra{0}a_{\bm p'}e^{-HT}\,\mathcal{H}\,e^{-HT}a^\dagger_{\bm p}\ket{0}$. In the case of solitons, $a$ and $a^\dagger$ will be replaced by $O^\dagger$ and $O$ respectively, but everything else is the same.

Since the classical radius is defined as the second moment of the energy density, \eqref{firsteq}, to define a quantum analogue, we need to consider the form factor $\bra{p'}\mathcal{H}(0)\ket{p}$. Taking derivatives of this w.r.t. to the momentum transfer $p'-p$ gives moments of the energy density; 
in particular, taking two derivatives gives the second moment, so we define the quantum radius by
\begin{align}R^2=M^{-2}\partial_{\ell^2}\bra{k}\mathcal{H}(0)\ket{k+\ell}\big{|}_{\ell=0}
\end{align}
which we will show gives $R^2=\frac{1}{2M^4} \bra{k}\mathcal{H}(0)K^2\ket{k}$ in two dimensions, where $K$ is the boost generator and $k^\mu$ is the rest momentum $(M,0)$. 
This is a matrix element between an initial and final soliton at rest. Therefore it can be directly evaluated by the methods of collective coordinate quantization, and the result can be organised in powers of the coupling constant. The leading term recovers the classical definition \eqref{firsteq}, and we will then go on to calculate  the first quantum correction.

The layout of the paper is as follows. In \S\ref{sec1} we discuss soliton operators and soliton states.
In \S\ref{sec:phi24} we introduce the relevant aspects of the classical $\phi_2^4$ model and its kink.
In \S\ref{sec:formfactor} we define the radius of the soliton, quantum mechanically. 
In \S\ref{sec:evaluation} we review the method of collective coordinate quantization and use it to compute the first quantum correction to the kink radius. 

A note on notation: in general spacetime dimensions, we will write spatial positions as $\bm x$ and spacetime positions as $x$ or $x^\mu$; in two dimensions, however, we write spatial positions as $x$ and spacetime positions as $x^\mu$.

\section{Soliton States and Operators \label{sec1}}

Consider a QFT of a real scalar, $\phi(\bm x)$, in $d+1$ dimensional Minkowksi space. 
The orthonormal basis states of the Hilbert space are $\ket{f}_{\phi}$ (where $f\equiv f(\bm x)$), which are eigenstates of $\phi(\bm x)$: $\phi(\bm x)\ket{f}_{\phi}=f(\bm x)\ket{f}_{\phi}$. Similarly, there is a dual orthonormal basis $\ket{g}_{\pi}$ of eigenstates of $\pi(\bm x)$, the conjugate momentum operator: $\pi(\bm x)\ket{g}_{\pi}=g(\bm x)\ket{g}_{\pi}$.

Now let the classical equations of motion possess a soliton solution. This means that in the rest frame of the soliton there exist initial data
$\phi_{cl}(\bm x)$ and $\pi_{cl}(\bm x)$
such that the energy density doesn't dissipate under classical time evolution. 
The soliton creation operator (first written down by Cahill \cite{Cahill1974}) is
\begin{equation}
    \mathcal{O}(\bm x)=\exp \left(-i\int \pi(\bm x+\bm a)\phi_{cl}(\bm a)d^d\bm a +i\int \phi(\bm x+\bm a) \pi_{cl}(\bm a)d^d\bm a \right), \label{mandel}
\end{equation}
and the annihilation (equivalently, anti-soliton) operator is $\mathcal{O}(\bm x)^\dagger$. 

As a consequence of the canonical commutation relations, the soliton operators cause a shift in the field value by the classical configurations,
\begin{align}
    \mathcal{O}(\bm x)^{-1}\phi(\bm y)\mathcal{O}(\bm x)&=\phi(\bm y)+\phi_{cl}(\bm y-\bm x), \\
     \mathcal{O}(\bm x)^{-1}\pi(\bm y)\mathcal{O}(\bm x)&=\pi(\bm y)+\pi_{cl}(\bm y-\bm x).
\end{align}
From this we deduce that 
$\mathcal{O}(\bm x)\ket{f(\bm y)}_{\phi}=\text{const.}\ket{f(\bm y)+\phi_{cl}(\bm y-\bm x)}_{\phi}$
i.e. a soliton centred on $\bm x$ has been added to the field configuration;\footnote{Alternatively, this follows from the fact that $\exp\left(-\int d^d\bm x\,g(\bm x)\pi(\bm x)\right)\ket{f}_{\phi}=\ket{f+g}_{\phi}$.} and a similar result applies in the dual basis.

$\mathcal{O}(\bm x)$ transforms as expected under a rotation $\bm R$. Assume that $\phi_{cl}(\bm x)$ and $\pi_{cl}(\bm x)$ are centered on $\bm x=0$ and are spherically symmetric. Then, denoting by $U$ the unitary operator acting on states,
\begin{equation}
U(\bm{R})\mathcal{O}(\bm x)U^{-1}(\bm{R})=\mathcal{O}(\bm{R}\bm x), \label{user1}
\end{equation}
and similarly
for a translation $\bm a$, 
\begin{equation}
U(\bm a)\mathcal{O}(\bm x)U^{-1}(\bm a)=\mathcal{O}(\bm x+\bm a). \label{user2}
\end{equation}
As mentioned in the introduction, an eigenstate of the Hamiltonian can be constructed by preparing a state at $t=0$ and time translating it to past Euclidean infinity.
The state at $t=0$ can be thought of as a proxy to the exact state. Because time translation commutes with $P_i$ and $J_{ij}$ (the spatial-momentum and angular-momentum generators), the proxy state must already have good spatial-momentum and angular-momentum quantum numbers. Further, the exact state (if it contains several particles) inherits the exchange statistics of the proxy state. 
In conventional field theory (where one is interested in elementary particles), proxy states are Fock states.
Given a suitable proxy, $\ket{n,\sim}$, the exact state $\ket{n}$ is,\footnote{This follows from inserting a complete set of energy eigenstates and picking out the state with lowest energy and non-vanishing overlap with $\ket{n,\sim}$.}
\begin{equation}
    \ket{n}=\lim_{T\to \infty} e^{ET}e^{-HT}\ket{n,\sim}/\braket{n|n,\sim}, \label{doubt}
\end{equation}
where $E$ is the energy $H\ket{n}=E\ket{n}$.
\eqref{doubt} projects the proxy state onto the lowest-energy exact-state (with non-vanishing overlap with the proxy-state) with the spatial-momentum and angular-momentum quantum numbers of the proxy-state. The factor $\braket{n|n,\sim}^{-1}$ ensures $\ket{n}$ is normalized to unity.
We should note that \eqref{doubt} avoids Haag's theorem \cite{Streater1989} 
because $e^{ET}e^{-HT}$ is not unitary. 
The factor $e^{ET}$ amputates external legs.\footnote{For a simple demonstration of this, imagine $\ket{n,\sim}$ was a Fock state. Then 
$e^{ET}$ is equivalent to $e^{H_0 T}$ since $H_0$ is assumed to have the same spectrum as $H$ (\cite{Weinberg2005} Sec 3.1). 
Then the difference between $H$ and $H_0$ will contain the difference between the bare mass (contained in $H$) and the exact mass (contained in $H_0$). These interaction vertices will then completely remove radiative corrections.
}

Although \eqref{doubt} looks like it is in Euclidean signature, it is actually physical. As an example, consider $\phi^4$ theory in 3+1 dimensions. Let $\braket{m|n}$ be a scattering amplitude, with $\ket{n,\sim}=\prod_{i=1}^n a^\dagger_{\bm k_i}\ket{0}$ and $\ket{m,\sim}=\prod_{j=1}^m a^\dagger _{\bm p_j}\ket{0}$. Using \eqref{doubt}, we get \begin{equation}    \braket{m|n}=C\int [d\phi]e^{-S[\phi]} \int \prod_{i=1}^n  d^4x_i e^{i\bm k_i\bm x_i-E_{\bm k_i}x_i^0}\Box _{x_i}\phi(x_i)\int \prod_{j=1}^m  d^4y_j e^{-i\bm p_j\bm y_j+E_{\bm p_j}y_j^0}\Box_{y_j} \phi(y_j), \end{equation} where $C=\braket{n|n,\sim}^{-1}\braket{m,\sim|m}^{-1}$, $\Box_x\equiv (\partial_{x^0}^2+\bm\nabla_{\bm x}^2-m^2)$, $E_{\bm k}=\sqrt{\bm k^2+m^2}$ and $S=\int d^4x(\dot\phi^2+(\bm\nabla \phi)^2+m^2\phi^2)/2+\lambda\phi^4$. Up to the factor $C$,  this is the amputated, Euclidean Greens function $G_E(k_1,...,k_n;-p_1,...,-p_m)$ with $k_i=(iE_{\bm k_i},\bm k_i)$ and $p_j=(iE_{\bm p_j},\bm p_j)$. The net result is that the external momenta are in Lorentz signature and the loop-momenta (i.e. unfixed momenta) are in Euclidean signature, which is exactly the prescription used to calculate physical amplitudes. Further, the missing factors of $i$ in the Feynman rules only affect the amplitude by an overall factor of $i$ (\cite{Ramond} Sec. 4.8).

Here, we need to construct a 1-particle soliton state. It is enough to construct the rest-state, from which the moving states can be obtained by a boost. Let $k^\mu=(M,0,...,0)$ be the rest momentum and let $\ket{k,\sim}$ be the proxy to the rest-state. It is \cite{Cahill1974}
\begin{equation}
    \ket{k,\sim}=\int d^d\bm x\,\,\mathcal{O}(\bm x) \ket{\Omega,\sim },\label{4.71}
\end{equation}
with
\begin{equation}
\ket{\Omega,\sim }=\ket{\phi(\bm x)=\phi_0}_{\phi}, \label{1mansil1}
\end{equation}
where $\phi_0$ is the minimum of the classical potential.
$\ket{\Omega,\sim }$ is a proxy state to the true vacuum state, $\ket{\Omega}$, since it has zero spatial-momentum and angular-momentum quantum numbers.

From \eqref{user1}, \eqref{user2}, it follows that $\ket{k,\sim}$ has zero spatial and angular momentum.
The exact state $\ket{k}$ is then
\begin{equation}
    \ket{k}=\text{const.}\frac{\lim_{T\to \infty} e^{MT}e^{-HT}\ket{k,\sim}}{\braket{k|k,\sim}}, \label{newlabel}
\end{equation}
where $M$ is the pole-mass of the particle (i.e. renormalized and quantum corrected) and `$\text{const}.$' allows for the possibility of adopting some different convention for the normalization of the particle state. The moving state $\ket{p}$ is 
\begin{equation}
    \ket{p}=U(\Lambda)\ket{k}, \label{mover}
\end{equation}
where $\Lambda: k\to p$.
Note that this definition is self-consistent: if we choose a different boost $\Lambda':k\to p$, it will be related to $\Lambda$ by a rotation and rotations leave $\ket{k}$ unchanged.

It is instructive to understand 
why the soliton operators take the form \eqref{mandel} from a different point of view.
Consider the matrix element $\bra{\psi_{out}}e^{-2HT}\mathcal{O}(\bm x_0)\ket{\psi_{in}}$. Transforming to the path integral, this is equal to 
\begin{equation}
    \int [d\phi][d\pi]\psi_{out}^*[\phi]\,\psi_{in}[\phi]\,e^{S'[\phi,\pi]}, \label{outsider}
\end{equation}
where
\begin{align}
    S'=S-i\int \pi(-T,\bm x_0+\bm a)\phi_{cl}(\bm a)d^d\bm a +i\int \phi(-T,\bm x_0+\bm a) \pi_{cl}(\bm a)d^d\bm a
\end{align}
and
$S=\int_{-T}^T dt \int d^{d}\bm x \left(i\pi\dot\phi-\mathcal H(\pi,\phi)\right)$. 

If we evaluate \eqref{outsider} perturbatively, we obtain an infinite sum of tree diagrams because the `free' part of $S'$ contains terms linear in the fields. These tree diagrams actually sum up to the classical background \cite{Cheung2020,Riccardo2019,Riccardo2020} which is determined by the equations of motion of $S'$ ($\delta S'/\delta \phi=\delta S'/\delta \pi=0$):
\begin{align}
    i\,\dot\pi(x)&=i\,\delta(t+T)\pi_{cl}(\bm x-\bm x_0) -\frac{\partial \mathcal H}{\partial \phi(x)}, \label{hamil1}\\
    i\,\dot \phi(x)&=i\,\delta(t+T) \phi_{cl}(\bm x-\bm x_0)+\frac{\partial \mathcal H}{\partial \pi(x)}. \label{hamil2}
\end{align}
Note that the usual background $\phi=\pi=0$ is not a solution. To build a solution, consider initial data $\phi(\bm x)=\pi(\bm x)=0$ at a very early time slice. Then at $t=-T$ there is an abrupt jump to values $\phi_{cl}(\bm x-\bm x_0)$ and $\pi_{cl}(\bm x-\bm x_0)$ respectively and then there is standard evolution (i.e. governed by $S$). In other words, let $\phi_{cl}(t,\bm x)$ and $\pi_{cl}(t,\bm x)$ be the solutions to the \emph{unsourced} EOM (i.e. the EOM derived from $S$) such that $\phi_{cl}(-T,\bm x)=\phi_{cl}(\bm x)$ and $\pi_{cl}(-T,\bm x)=\pi_{cl}(\bm x)$ is an initial condition. Then the solution to \eqref{hamil1}, \eqref{hamil2} is 
\begin{align}
    \phi(x)&= \theta(t+T)\phi_{cl}(t,\bm x-\bm x_0),\\
    \pi(x)&=\theta(t+T)\pi_{cl}(t,\bm x-\bm x_0).
\end{align}
The step function 
indicates the creation of a soliton at $t=-T$. The operator \eqref{mandel} is the $1+1$ dimensional topological soliton analogue of t'Hooft's topology changing operators in $2+1$ dimensional gauge theory \cite{Hooft1978}.

\section{The classical \texorpdfstring{$\phi_2^4$}{Lg} model and the kink  \label{sec:phi24}} 
We will deal in this section with the purely classical aspects of the $\phi_2^4$ model,
a theory of one real scalar field $\phi$ in 1+1 dimensions. 
In the conventions of \cite{DHN1974} the Lagrangian is 
\begin{equation}
    \mathcal{L}=\frac{1}{2}(\partial_\mu \phi)^2-\frac{1}{4\lambda^{2}}\phi^4+\frac{1}{2}m^2\phi^2-\frac{m^4\lambda^2}{4}.
\end{equation}
With $m^2>0$, the potential is a Mexican hat with two absolute minima of zero at $\phi=\pm \lambda m$.   
We will follow \cite{GervaisPert1975} to set $m=1$ and work instead with the Lagrangian 
\begin{equation}
    \mathcal{L}=\frac{1}{2}(\partial_\mu \phi)^2-\frac{1}{4\lambda^{2}}\phi^4+\frac{1}{2}\phi^2-\frac{\lambda^2}{4}.\label{eqn:phi4Lagrangian}
\end{equation}
It is easy to restore factors of $m$ at the end using the fact that the mass dimensions of $\phi$ and $\lambda$ are $0$ and $-1$ respectively.
The equation of motion (EOM) which follows is
\begin{equation}
    \Box \phi+\lambda^{-2}\phi^3-\phi=0.  \label{EOM}
\end{equation}
This admits a static soliton solution:
\begin{equation}
    \phi_{cl}(x)=\lambda\tanh(x/\sqrt 2).
\end{equation}
The solution interpolates from one vacuum value $-\lambda$ at $x=-\infty$ to the other $\lambda$ at $x=\infty$. The energy of a static solution is $\int dx~ (\phi'^2/2+V(\phi))$ (where $V=\phi^4/(4\lambda^2)-\phi^2/2+\lambda^4/4$). However,  a virial relation 
\cite{Goldstone1975}  ensures that $\int dx ~V=\int dx~ \phi'^2/2$. 
Therefore the energy is $\int dx~\phi'^2$. This gives us the classical mass of the soliton: 
\begin{equation}
    M_0=2\sqrt 2\lambda^2/3. \label{classicalmass}
\end{equation} 
We also have 
\begin{equation}
    \int_{\infty}^\infty dx\,\phi_{cl}'(x)\phi_{cl}(x)=\phi_{cl}^2\big|_{-\infty}^\infty=0.\label{mightneed}
\end{equation}
To establish the spectrum of perturbations, we expand around $\phi_{cl}$: $\phi=\phi_{cl}+\Psi$. Using this in the EOM \eqref{EOM}, the perturbation satisfies
\begin{equation}
    \Box\Psi+\left(3\tanh^2\left(x/\sqrt 2\right)-1\right)\Psi=0, \label{preschro}
\end{equation}
to linear order.
Letting $\Psi(t,x)=\exp(i \omega t)\psi(x)$, \eqref{preschro} becomes the Schr\"odinger equation 
\begin{equation}
    \psi''+\left(\omega^2+1-3\tanh^2\left(x/\sqrt 2\right)\right)\psi=0.
\end{equation}
The spectrum $\mathcal S$ consists of two discrete eigenvalues with $\omega=0$, which is the translational zero mode, and $\omega=\sqrt{3/2}$, which is localised,  plus a continuum  labelled by $k\in\mathbb{R}$ with $\omega_k=\sqrt{k^2+2}$. 
The eigenfunctions are
\begin{align}
    \psi_b&=\frac{\phi_{cl}'}{\sqrt{M_0}}\hspace{1cm} \omega_b=0,\\
    \psi_a&=\sqrt\frac{3}{2\sqrt{2}}\,\frac{\sinh (x/\sqrt 2)}{\cosh^2(x/\sqrt 2)} \hspace{1cm} \omega_a=\sqrt{3/2},\\
    \psi_k&=N_k^{-1}\exp(ikx)\left(3\tanh^2(x/\sqrt 2) -3\sqrt 2ik\tanh (x/\sqrt 2) -1-2k^2\right),\label{eqn:psikdef}\\
    \omega_k&=\sqrt{k^2+2}, \quad N_k=\sqrt{2(k^2+2)(2k^2+1)}.
\end{align}
Note that $\psi_{k}^*=\psi_{-k}$. These functions are orthonormal with the measure $dx$ and satisfy the completeness relation
\begin{align}
    \delta(x-x')=\psi_b(x)\psi_b(x')+\psi_a(x)\psi_a(x')+\int \frac{dk}{2\pi} \psi^*_k(x')\psi_k(x). \label{eqn:complete}
\end{align}
\noindent We record for later the useful relationship 
\begin{align}
    \psi_k(x)\psi_k^*(x')&=e^{ik(x-x')}\left(1+N_k^{-2}\sum_{l=0}^3 k^l\,{A}_l(x,x')\right),
    \label{use1}
\end{align}
where
\begin{align}
    {A}_3(x,x')&= 6i\sqrt 2(\tau-\tau'), \label{1y}\\
    {A}_2(x,x')&= -6\tau^2-6\tau'^2+18\tau\tau'-6, \label{2y}\\
    {A}_1(x,x')&= 3i\sqrt2 (\tau-\tau')(3\tau\tau'+1),\label{3y}\\
    A_0(x,x')&= (3\tau^2-1)(3\tau'^2-1)-4, \label{4y}
\end{align}
with $\tau\equiv\tanh(x/\sqrt 2)$ and $\tau'\equiv\tanh(x'/\sqrt 2)$.

Of course \eqref{EOM} also has the trivial static solution
\begin{equation}
    \phi(x)=\lambda.
\end{equation}
The spectrum of fluctuations about this solution is labelled by $k\in\mathbb R$, with orthonormal functions
\begin{equation}
    \psi_{k}^0(x)=e^{ikx},
\end{equation}
and completeness relation
\begin{align}
    \delta(x-x')=\int \frac{dk}{2\pi} \psi^{0*}_k(x')\psi^0_k(x).
\end{align}

\section{Form Factors and the Radius \label{sec:formfactor}}
In this section we relate the radius of the soliton of physical mass $M$ to a form factor, in two dimensions. 

For a particle state $\ket{p^\mu}$, a form factor is $\bra{p'}\mathcal{O}\ket{p}$, for some operator $\mathcal{O}$. A form factor, when expanded in powers of the momentum transfer, $p'-p$, reveals information about the internal structure of the particle in question, and with different operators, $\mathcal{O}$, one can probe different aspects of the internal structure. 
As discussed in the introduction, our goal will be to calculate the radius of the soliton by computing the `width' of the energy density, i.e. by computing $\bra{p'}\mathcal H\ket{p}$, where $\ket{p}$ is the soliton state and $\mathcal{H}$
is the energy density.

We use the single-particle normalization conditions 
\begin{align}
\braket{k|k'}&=4\pi k^0\,\delta(k-k'),\label{resc}\\
U(\Lambda)\ket{k}&=\ket{\Lambda k}, \label{wein}
\end{align}
so that the mass dimension of $\ket{k}$ is $0$. Consider
\begin{equation}
    \Gamma^{\mu\nu}(k,k'):=\bra{k}T^{\mu\nu}(0,0)\ket{k'},
\end{equation}
where $T^{\mu\nu}(t,x)$ is the energy-momentum tensor.
The mass dimension of $\Gamma$ is $+2$.
A standard set of manipulations 
\cite{Weinberg2005} (Sec. 10.6) shows that
\begin{align}
\Gamma^{\mu\nu}(k,k')&=\frac{1}{2}\left(F_1(l^2)(k+k')^\mu (k+k')^\nu
+F_5(l^2)\left(\eta^{\mu\nu}-l^{-2}l^\mu l^\nu\right)\right),\label{rahm}
\end{align}
where $l^\mu\equiv (k-k')^\mu$. The form factors $F_1$ and $F_5$, which encode everything non-trivial about $\Gamma^{\mu\nu}$, are undetermined, real functions of $l^2$ and $M$, 
and satisfy $F_1(0)=1$ and $F_5(0)=0$.

The radius of the particle is related to the first derivatives of the form factors, $F_1'(0)$ and $F_5'(0)$.
The form factors depend only on the momentum difference, so we can conveniently choose $k^\mu=(M,0)$, $k'^\mu=(\sqrt{M^2+q^2},q)$ and expand in powers of $q$. The momentum transfer satisfies $l^2=-q^2+{O}(q^4)$. 
We define the radius, $R$, by
\begin{equation}
    R^2=\frac{1}{M^2}\partial_{-q^2}\Gamma^{00}(k,k')\big{|}_{q=0}. \label{radiusdef}
\end{equation}
As we will show later, \eqref{leading},  to leading order in $\hbar$ (i.e. classically), this definition gives $R^2=\int dx\,x^2 \mathcal H_{cl}(x)/\int dx\,\mathcal{H}_{cl}(x)$ which coincides with the discussion in \S\ref{bcconj}.
Now, let $\Lambda$ be the Lorentz transformation that takes $k\to k'$; 
the corresponding operator $U(\Lambda)$ expanded in powers of $q$ is,\footnote{This follows from $U=\exp(i\omega_{10}K)$ where $\omega_{\mu\nu}=\eta_{\mu\rho}\omega^{\rho}_\nu$ and $\omega^{\mu}_\nu$ is determined from $\Lambda^\mu_\nu=\delta^\mu_\nu+\omega^\mu_\nu+O(\omega^2)$ -- which gives $\omega_{10}=-q/M$.} 
\begin{equation}
    U(\Lambda)=1-\frac{iq}{M}K-\frac{q^2}{2M^2}K^2+{O}(q^3),\label{bew}
\end{equation}
where $K$ is the boost generator,
\begin{equation}
    K=\int dx ~x\,T^{00}(0,x). \label{explic}
\end{equation}
Using \eqref{wein}, \eqref{bew} in \eqref{radiusdef} shows that
\begin{equation}
     R^2=\frac{1}{2M^4}\bra{k}\mathcal{H}(0,0)K^2\ket{k},\label{raddef}
\end{equation}
where the energy density $\mathcal{H}(t,x)=T^{00}(t,x)$,  
and then using \eqref{explic} gives 
\begin{equation}
    R^2= \bra{k} O \ket{k},
    \label{rsq1}
\end{equation}
where
\begin{equation}
    O\equiv\frac{1}{2M^4}\int x\, x'\,\mathcal{H}(0,0)\mathcal{H}(0,x)\mathcal{H}(0,x')dx\,dx'.
\end{equation} 

Taking this definition of the radius and applying it to the soliton case using 
 \eqref{4.71}, \eqref{newlabel}
 we then have,\footnote{The denominator of \eqref{label1} follows from applying \eqref{4.71}, \eqref{newlabel} to $\braket{k|k}$ and eliminating the common factor of $|\braket{k|k,\sim}|^2$ between this and $R^2$.}
\begin{equation}
    R^2=\frac{\lim_{T\to \infty}e^{2MT} \int\bra{\Omega,\sim}\mathcal{O}^\dagger(y)e^{-HT}O e^{-HT}\mathcal{O}( z)\ket{\Omega,\sim}dz\,dy}{\braket{k|k}^{-1}\lim_{T\to \infty}e^{2MT}\int \bra{\Omega,\sim}\mathcal{O}^\dagger(y)e^{-2HT}\mathcal{O}(z)\ket{\Omega,\sim}dz\,dy}. \label{label1}
\end{equation}
For the moment, we will suppose that the system is in a spatial box of size $L$.
The factor of $\braket{k|k}$ in \eqref{label1} is fixed since \eqref{resc} reads
$\braket{k|k}=2ML$ in a box of size $L$. 
Now, since $\pi_{cl}=0$, the soliton operator is $\mathcal{O}(x)=\exp(-i\int da~\phi_{cl}(a)\pi(x+a))$. Using 
$\ket{\Omega,\sim}=\ket{0}_{\phi}$ we then have $\mathcal{O}(x)\ket{\Omega,\sim}=\ket{\phi_{cl}(\cdot -x)}_{\phi}$ (this is the configuration of a soliton centered on $x$). For convenience we will denote it as $\ket{x}\equiv \ket{\phi_{cl}(\cdot -x)}_{\phi}$.
Therefore
\begin{equation}
    R^2=\frac{\lim_{T\to \infty}e^{2MT} \int\bra{y}e^{-HT}O e^{-HT}\ket{z}dz\,dy}{\frac{1}{2ML}\lim_{T\to \infty}e^{2MT}\int \bra{y}e^{-2HT}\ket{z}dz\,dy}. \label{rint}
\end{equation}

\section{Evaluation of the radius for the \texorpdfstring{$\phi_2^4$}{Lg} kink}\label{sec:evaluation}

The Hamiltonian of the $1+1$ dimensional $\phi^4$ model in the conventions of \eqref{eqn:phi4Lagrangian} is
\begin{align}
    H(\pi,\phi)&=\int \mathcal{H}(x) \,dx\nn\\
    &= \int  \frac{1}{2}\left(\pi^2+\phi'^2-(1+\delta_{m^2})\phi^2+\frac{\lambda^2}{2}(1+\delta_{m^2})^2+\frac{\phi^4}{2\lambda^2}+2C(x)\right) dx, \label{hamdens}
\end{align}
where $ \delta_{m^2}$ is the mass counterterm, and $C(x)$ is the vacuum energy density counterterm chosen so that $H\ket{\Omega}=0$ and formally given by
\begin{equation}
    C(x)=-\bra\Omega\frac{1}{2}\left(\pi^2+\phi'^2-(1+\delta_{m^2})\phi^2+\frac{\lambda^2}{2}(1+\delta_{m^2})^2+\frac{\phi^4}{2\lambda^2}\right)\ket\Omega. \label{C}
\end{equation}
Note that  the renormalized mass parameter is $m=1$ 
(as observed in \S\ref{sec:phi24} factors of $m$ can be restored by dimensional analysis). In this convention, the Feynman rules are given in  \cite{GervaisPert1975}.

To evaluate $R^2$ from \eqref{rint} requires several steps: the first is to quantize using the collective coordinate formalism;  then we expand to obtain an expression equivalent  to \eqref{rint} to first sub-leading order in $\lambda^{-1}$; 
and we complete the calculation using point-splitting regularization.

\subsection{Collective Coordinate Quantization \label{CCQ}}

$R^2$, given in \eqref{rint}, involves matrix elements between initial and final states in which there is a single soliton at rest. The method of computing such matrix elements is collective-coordinate-quantization (CCQ) \cite{GervaisPoint1976}.

CCQ consists of a change of variables from the usual field variables, to a soliton position variable and a perturbation field around the soliton:
$\phi(x)\to(X,\chi(x))$. The domains of the new variables are $\chi(x)\in \mathbb{R}$ and $X\in\mathbb{R}$.
$X(\phi)$ is the soliton position and it is defined by
\begin{equation}
    \int_{-\infty}^{\infty} \phi_{cl}'(x)\phi(x+X(\phi))\,dx=0. \label{defsec}
\end{equation}
For example, if $\phi(x)=\phi_{cl}(x-x_0)$,
then $X(\phi)=x_0$ due to \eqref{mightneed}.
There is no proof that \eqref{defsec} has only one solution for each $\phi$, but this is a global issue which will not be important when we work in perturbation theory \cite{Callan1975} (p. 36). 
$\chi(x)$ is defined as
\begin{equation}
    \chi(x)\equiv\phi(x+X(\phi))-\phi_{cl}(x). \label{saver}
\end{equation}
The inverse of this relation is
\begin{equation}
    \phi(x)=\phi_{cl}(x-X)+\chi(x-X). \label{save1}
\end{equation}
As a consequence of \eqref{mightneed}, \eqref{defsec}, $\chi$ satisfies the constraint
\begin{equation}
   Q[\chi]=0, \label{constraint1}
\end{equation}
where 
\begin{equation}
   Q[\phi]\equiv\int_{-\infty}^{\infty} M_0^{-1/2}\phi_{cl}'(x)\phi(x)\,dx.
\end{equation}
It can be checked that this change of variables is a bijection. It can be shown that the Jacobian of the transformation is
\begin{equation}
    J=\frac{\sqrt{M_0}}{\int dx (\phi_{cl}'+\chi')\phi_{cl}'}. \label{jack}
\end{equation}
We define $P$, dual to $X$, such that $[P,X]=-i$; and 
$\bar \pi(x)$ dual to $\chi(x)$ satisfying \cite{Tomboulis1975}
\begin{equation}
    Q[\bar \pi]=0, \label{constraint2}
\end{equation}
and
\begin{equation}
    i[\bar\pi(x),\chi(y)]=\delta(x-y)-\frac{\phi_{cl}'(x)\phi_{cl}'(y)}{M_0}.
\end{equation}
There is the following relationship between the old and new variables \cite{Tomboulis1975}:
\begin{align}
    \pi(x)=\bar\pi(x-X)-\frac{1}{2}\left\{\frac{\phi_{cl}'(x-X)}{\int \phi_{cl}'(\chi'+\phi_{cl}')},P+\int \bar\pi\chi' \right\},
    \label{finally}
\end{align}
where $\{,\}$ is the anti-commutator.

By substituting $\pi(x)$ and $\phi(x)$, from \eqref{finally} and \eqref{save1} respectively, into the Hamiltonian $H(\phi,\pi)$, we obtain the dual Hamiltonian, $H(X,\chi,P,\bar\pi)$. (Abusing notation we have denoted the dual Hamiltonian with the same symbol.)

We can also construct path integrals in the new variables. Consider a matrix element, $\bra{\psi_{out}}e^{-HT}{O}e^{-HT}\ket{\psi_{in}}$, where  $O$ is an operator at time zero that is function of $X,\chi(x),P,\bar\pi(x)$.
It can be written as a path integral:
\begin{align}
    \bra{\psi_{out}}e^{-HT}{O}e^{-HT}\ket{\psi_{in}}&= \int [dX] [d\chi] [dP] [d\bar\pi] \,\psi_{in}\left(X,\chi\right)\psi^*_{out}\left(X,\chi\right)\delta\left(Q[\chi]\right)\,\delta\left(Q[\bar\pi]\right)\nonumber\\
    \times O(P,\bar\pi,X,\chi)&\exp\int_{-T}^T dt\left( i P\dot X+i\int dx (\bar\pi \dot\chi)-H(X,\chi,P,\bar\pi)\right), \label{Rahbro}
\end{align}
where the integral is over configurations \begin{align} X(t),\chi(t,x),P(t),\bar\pi(t,x): t\in[-T,T], x\in[-\infty,\infty]. 
\end{align}
The delta-functions arise due to \eqref{constraint1}, \eqref{constraint2}, and enforce $Q[\chi]=Q[\bar\pi]=0$, $\forall\,t$. The time argument of the fields in \eqref{Rahbro} have been suppressed: in $O$, all fields have time argument $t=0$ and in $\psi_{in/out}$ they are $t=\mp\infty$ respectively.
Also in \eqref{Rahbro}, we have defined the wavefunctions $\psi_{in/out}(X,\chi)=\braket{X,\chi|\psi_{in/out}}$, where $\ket{X,\chi}$ are the simultaneous eigenstates of the operators $X$ and $\chi(x)$, normalized such that $1=\int dX[d\chi]\, \ket{X,\chi}\bra{X,\chi}$ (where the integrals are over configurations $X,\chi(x)$ at fixed time). For example, if $\ket{\psi}=\ket{\phi_{cl}(\cdot-x_0)}_\phi$, then $\psi=\sqrt J\,\delta(X-x_0)\delta(\chi)=M_0^{-1/2}\delta(X-x_0)\delta(\chi)$ using \eqref{jack}.

\subsection{Radius to \texorpdfstring{${\mathcal O}(\lambda^{-2})$}{Lg}}

In this section we evaluate $R^2$ in powers of $\lambda^{-1}$, determining the leading and subleading terms. (We should note that the expansion in powers of $\lambda^{-1}$ is equivalent to the expansion in powers of $\hbar$ (\cite{korepin} Sec. 1.1), so what we are working out is the first quantum correction to $R$.)

We begin by expanding \eqref{finally}:\footnote{The order of $\lambda$ can be counted by recalling $\phi_{cl}\propto \lambda$, $M_0\propto \lambda^2$, and $\delta_{m^2}\propto\lambda^{-2}$.}
\begin{equation}
    \pi(x)=\bar\pi(x-X)-\frac{1}{2M_0}\left\{\phi_{cl}'(x-X),P+\int \chi'\bar\pi \right\}+O(\lambda^{-2}). \label{sub1}
\end{equation} 
The energy density, from \eqref{hamdens}, is 
\begin{equation}
    \mathcal{H}(x)= \frac{1}{2}\left(\pi(x)^2+\phi'(x)^2-(1+\delta_{m^2})\phi(x)^2+\frac{\phi^4(x)}{2\lambda^2}+\frac{\lambda^2}{2}(1+\delta_{m^2})^2+2C(x)\right).
\end{equation}
Substituting \eqref{sub1}, \eqref{save1} into this, we get $\mathcal{H}(x)=\tilde{\mathcal{H}}(x-X)$, where
\begin{align}
    \tilde{\mathcal{H}}(x)&=a(x) +b(x)+c(x)+d(x)+{O}(\lambda^{-2}),\label{eqn:Htilde}
\end{align}
where $a, b, c, d$ are ${\mathcal O}(\lambda^2, \lambda,1,\lambda^{-1})$ respectively and are given by
\begin{align}
    a(x)&=\frac{1}{2}\left(\phi_{cl}'^2-\phi_{cl}^2+\frac{\phi_{cl}^4}{2\lambda^2}+\frac{\lambda^2}{2}\right),\nn\\
    b(x)&=(\phi_{cl}'\chi)',\nn\\
    c(x)&= \frac{1}{2}\left(\bar\pi^2+\chi'^2-\chi^2+\frac{3}{\lambda^2}\phi_{cl}^2\chi^2-\delta_{m^2}(\phi_{cl}^2-\lambda^2) \right)+C_0(x), \nn\\
    d(x)&=-\frac{1}{4 M_0}\left\{\bar\pi,\left\{\phi_{cl}',P+\int \chi'(y)\bar\pi(y)\, dy \right\}\right\}-\delta_{m^2}\phi_{cl}\chi+\frac{\chi^3\phi_{cl}}{\lambda^2}+C_{-1}(x).\label{eqn:cx}
\end{align}
Here $C_0(x)$ and $C_{-1}(x)$ are the ${ O}( \lambda^0)$ and ${ O} (\lambda^{-1})$ components respectively  of $C(x)$.
Integrating $\tilde {\mathcal H}(x-X)$ w.r.t. $x$, gives 
\begin{align}
    H=M_0+H_{free}+H_{int}+O(\lambda^{-2})
\end{align}
where \cite{GervaisPert1975}
\begin{align}
    H_{free}&=\int dx ~\frac{1}{2}\left(\bar\pi^2+\chi'^2-\chi^2+\frac{3}{\lambda^2}\phi_{cl}^2\chi^2-\delta_{m^2}(\phi_{cl}^2-\lambda^2) \right)+\int dx\, C_0(x),\\
    H_{int}&=\int dx \left(-\delta_{m^2}\phi_{cl}\chi+\frac{\chi^3\phi_{cl}}{\lambda^2}\right)+\int dx \,C_{-1}(x),
\end{align}
where we have used the constraint \eqref{constraint2} and the fact that integration variables can be shifted by $X$ provided that every factor under the integral commutes with $X$.

The numerator and denominator of $R^2$ \eqref{rint} is of the form of the matrix element, $\bra{\psi_{out}}e^{-HT}{O}e^{-HT}\ket{\psi_{in}}$, of the previous section, with $\ket{\psi_{in}}=\ket{z}$, $\ket{\psi_{out}}=\ket{y}$. The wavefunctions of the `in' and `out' states are then (as remarked in the previous section) $M_0^{-1/2}\delta(X-z)\delta(\chi)$ and $M_0^{-1/2}\delta(X-y)\delta(\chi)$ respectively. 
Applying \eqref{Rahbro} we thus have,\footnote{To obtain this formula we have made use of the fact that the $\delta(\chi(t=\pm\infty))$ factors affect the amplitude by only an overall constant factor. 
}
\begin{align}
    &\int\bra{y}e^{-HT}\mathcal{H}(x_1)\mathcal{H}(x_2)\mathcal{H}(x_3) e^{-HT}\ket{x}dx\,dy \nn\\
    &= \xi\int [dX] [d\chi] [dP] [d\bar\pi] \nn\\
    &\quad\quad\times\tilde{\mathcal{H}}(x_1-X)\tilde{\mathcal{H}}(x_2-X)\tilde{\mathcal{H}}(x_3-X)\exp\int_{-T}^T dt\left( i P\dot X+i\int dx (\bar\pi \dot\chi)-H\right), \label{prun}
\end{align}
where $\xi$ is an overall constant.

To $ O (\lambda^{-2})$ 
the highest order surviving term 
in $\tilde{ \mathcal{H}}(x)$ is $c(x)$, and  in $H$ is $H_{int}$. Then, to the order of our calculation, $H$ or $\tilde{\mathcal{H}}$ doesn't depend on $P$; so the integral over $P$ gives $\delta(\dot X)$. The integral over $X$ is then restricted to constant paths $X(t)=x_0$, giving
\begin{align}
    \xi\int  [d\chi]  [d\bar\pi] \int dx_0 ~\tilde{\mathcal{H}}(x_1-x_0)\tilde{\mathcal{H}}(x_2-x_0)\tilde{\mathcal{H}}(x_3-x_0)~\exp\int_{-T}^T dt\left(i\int dx (\bar\pi \dot\chi)-H\right).
\end{align}
Similarly we have
\begin{align}
    \int \bra{y}e^{-2HT}\ket{x}dx\,dy=\xi\,L\int  [d\chi]  [d\bar\pi] ~\exp\int_{-T}^T dt\left( i\int dx (\bar\pi \dot\chi)-H\right),
\end{align}
where the factor of $L$ arises from the $x_0$ integral. Using these results in \eqref{rint} gives
\begin{align}
     R^2&=\frac{
     \int [d\chi][d\bar\pi] e^{S[\bar\pi,\chi]}
     \int  dx\,dx'\,dx''\left(x-x''\right)\left(x'-x''\right)\, \tilde{\mathcal{H}}(x'')\,\tilde{\mathcal{H}}(x)\,\tilde{\mathcal{H}}(x')}{M^3\int [d\chi][d\bar\pi]e^{S[\bar\pi,\chi]}},
     \label{path2}
\end{align}
where 
$S=S_{free}+S_{int}$ and
\begin{align}
    &S_{free}= \int d^2x\left(i\bar\pi\dot\chi-\frac{1}{2}\bar\pi^2-\frac{1}{2}\chi'^2+\frac{1}{2}\chi^2-\frac{3}{2\lambda^2}\phi_{cl}^2\chi^2\right), \label{sfree}\\
    &S_{int}=\int d^2x \left(-\frac{\phi_{cl}\chi^3}{\lambda^2}+\delta_{m^2}\phi_{cl}\chi\right). \label{a6ref}
\end{align}
Dropping everything beyond the first sub-leading order, 
we obtain
\begin{align}
    R^2&=M^{-3}(\eta+\gamma +{O}(\lambda^3)) 
    \label{curlym}
\end{align}
where $\eta, \gamma$ are ${O}(\lambda^6)$ and ${O}(\lambda^4)$ respectively and are given by
\begin{align}    
    \eta&=\int  a_1a_2a_3\,d\mu, \label{curlym2}\\
    \gamma&=\int d\mu \big( a_1a_2\langle S_{int}\,b_3\rangle
    +a_1\langle b_2 b_3\rangle
    +\langle c_1\rangle a_2a_3
    +\text{cyclic permutations of {1,2,3}}\big).~~~~\label{curlym1}
\end{align}
Here $a_{1,2,3}$ stand for $a(x''),a(x),a(x')$ respectively, and similarly for $b_{1,2,3}$ and $c_{1,2,3}$.
The integration measure is $ d\mu\equiv  dx\,dx'\,dx'' (x-x'')(x'-x'')$, and $\langle\cdot\rangle$ denotes expectation with respect to $S_{free}$.

In fact $a(x)$ is the classical energy density, so $\int  a(x)dx=M_0=2\sqrt 2\lambda^2/3$, $\int x\,a(x)dx=0$ and the leading term is given by 
\begin{equation}
    \eta=M_0^2 a_{(2)},\label{leading}
\end{equation}
where 
\begin{equation}
a_{(2)}\equiv \int x^2 a(x)=\frac{\sqrt{2}m\lambda^2}{9}  \left(\pi ^2-6\right).
\end{equation}
This leading contribution to $R^2$ \eqref{curlym} is simply the classical radius squared $R_{cl}^2$ \eqref{firsteq}.

The quantum correction simplifies to 
\begin{equation}
    \gamma=c_{(2)}M_0^2+2c_{(0)}M_0\,a_{(2)}-2M_0^2\,\int \langle S_{int}\,\chi(x)\rangle\, \phi_{cl}'(x)\, x\,dx \label{eqn:gamma}
\end{equation}
where $c_{(n)}\equiv \int x^n \langle c(x)\rangle dx$ 
and we have used
\begin{enumerate}
    \item \label{refa} $\int b  =\int (\phi_{cl}'\chi)'=0$ because $\phi_{cl}'$ is exponentially damped at large $|x|$,
    \item $\int x \,b  =\int x (\phi_{cl}'\chi)'=[x\phi_{cl}'\chi]^{\infty}_{-\infty}+\int \phi_{cl}'\chi =0$ because of the constraint on $\chi$ (see \eqref{constraint1}).
\end{enumerate}

A similar, though less involved, calculation gives the
 quantum corrected mass of the kink. This follows from $M=\braket{k|k}^{-1}\bra{k}H\ket{k}$ and using all the steps that led from \eqref{raddef} to \eqref{curlym}-\eqref{curlym1}, which gives
\begin{align}
    M&=\int_{-\infty}^\infty dx(a(x)+\langle c(x)\rangle )+{O}(\lambda^{-1})=M_0+c_{(0)}+{O}(\lambda^{-1}).  \label{eqn:mass}
\end{align}

\subsection{Free field expectation values\label{subsec:freefield}}

As usual, expectation values of polynomials of $\bar\pi$, $\chi$ w.r.t. $S_{free}$ can be constructed from the two-point functions, which are \cite{GervaisPert1975}:
\begin{align}
    \langle\chi(t,x)\chi(t',x')\rangle&=G(t-t';x,x'), \\
    \langle\chi(t,x)\bar\pi(t',x')\rangle&=i\partial_{t'}G(t-t';x,x'),\\
    \langle\bar\pi(t,x)\bar\pi(t',x')\rangle&=-\partial_{t}\partial_{t'}G(t-t';x,x')+\Delta(t-t';x,x'),
\end{align}
with
\begin{align}
    G(t-t';x,x')&=\int \frac{d\omega}{2\pi}e^{i\omega(t-t')}\left(\frac{\psi_a(x)\psi_a^*(x')}{\omega^2+\omega_a^2}+\int \frac{dk}{2\pi}\frac{\psi_k(x)\psi_k^*(x')}{\omega^2+k^2+2}\right), \label{prop1}\\
    \Delta(t-t';x,x')&=\delta(t-t')\left(\psi_a(x)\psi_a^*(x')+\int\frac{dk}{2\pi}\psi_k(x)\psi_k^*(x')\right). \label{prop2}
\end{align}
To obtain $C_0(x)$ ($C_{-1}(x)$ makes no contribution to the calculation of $R^2$ at one loop order), we use $\ket{\Omega}=\braket{\Omega|\lambda}_{\phi}^{-1}\lim_{T\to\infty}e^{-HT}\ket{\lambda}_{\phi}$ in \eqref{C}, and expand about the vacuum state by setting  $\phi(t,x)=\lambda+\varphi(t,x)$ in the path integral. 
This gives
\begin{align}
     C_0(x)&=-\frac{1}{2}\frac{\int [d\varphi] [d\pi] \left(\pi(x)\pi(x')+\left(\partial_x\partial_{x'}+2\right)\varphi(x)\varphi(x')\right)e^{S'_{free}}}{\int [d\varphi] [d\pi] e^{S'_{free}}}, \label{eqn:C0}
\end{align}
with $S'_{free}=\int dx dt \left(i\pi\dot\varphi-\pi^2/2-\varphi'^2/2-\varphi^2\right)$. This can be evaluated using  
\begin{align}
    \langle\varphi(t,x)\varphi(t',x')\rangle'&= G^0(t-t';x,x'), \label{eqn:G0}\\
    \langle\varphi(t,x)\pi(t',x')\rangle'&=i\partial_{t'} G^0(t-t';x,x'),\\
    \langle\pi(t,x)\pi(t',x')\rangle'&=-\partial_{t}\partial_{t'} G^0(t-t';x,x')+ \Delta^0(t-t';x,x'),
\end{align}
where $\langle\cdot\rangle'$ is expectation w.r.t. $S_{free}'$, and
\begin{align}
     G^0(t-t';x,x')&=\int \frac{d\omega\,dk}{4\pi^2}\frac{e^{i\omega(t-t')}e^{ik(x-x')}}{\omega^2+k^2+2}, \label{compare1}\\
    \Delta^0(t-t';x,x')&=\delta(t-t')\int \frac{dk}{2\pi}e^{ik(x-x')}. \label{compare2}
\end{align}
Finally it is useful to define
\begin{align}
    G^s(t-t';x,x')&= G(t-t';x,x')- G^0(t-t';x,x')\nn\\
    &=\int \frac{d\omega}{2\pi}e^{i\omega(t-t')}\frac{\psi_a(x)\psi_a^*(x')}{\omega^2+\frac{3}{2}}
    \nn\\
    &\quad+\int \frac{d\omega\,dk }{4\pi^2 }\frac{e^{i\omega(t-t')}e^{ik(x-x')}}{\omega^2+k^2+2}\frac{\sum_{l=0}^3 k^l\,{A}_l(x,x')}{2(k^2+2)(2k^2+1)},\label{eqn:GsAppB}
    \\
    \Delta^s(t-t';x,x')&=\Delta(t-t';x,x')-\Delta^0(t-t';x,x')\nn\\
    &=-\delta(t-t')\,\psi_b(x)\psi_b(x').\label{eqn:DsAppB}
\end{align}

\subsection{Regularization}

To complete the evaluation of $R^2$ 
to one loop order  requires the regularization
of divergences both for the mass counterterm and for the vacuum energy subtraction. 
In previous calculations of the mass shift \cite{DHN1974}, this was done by working at finite volume with a momentum cut-off and taking the volume and cut-off to infinity at the end of the calculation. 
However here we need the density distributions, not just their spatial integrals, and it is more convenient to use point-splitting regularisation.  
Divergences associated with products of two fields at the same space-time points are regularised by point-splitting field products $\Phi(0,x)^2$ to $\Phi(0,x)\,\Phi(t,x')$ and taking the coincidence limit, $t\to 0,\,x'\to x$, at the end of the calculation once all the divergences have cancelled. In this scheme the mass counter-term, which is computed in the trivial vacuum, is given by
\eqn{\delta_{m^2}=\frac{3}{\lambda^2}G^0\argsO,\label{eqn:deltam}}
where $G^0$ is the propagator for $\phi$ computed in the vacuum and is given in \eqref{eqn:G0}.

To compute $\langle c(x)\rangle$ \eqref{eqn:cx}, 
point-splitting the products and performing the functional integral gives
\AMSeqn{
c_{ps}(x,x') =\left(c^1_{ps}\args-c^0_{ps}\args\right)\vert_{t=0}\,,\label{eqn:cps}}
where, from \eqref{eqn:cx} and \eqref{eqn:C0},
\AMSeqn{c^1_{ps}\args&=\half(\partial_t^2+\partial_x\partial_{x'}+\frac{3}{\lambda^2}\phicl{2}{x}-1)\,G\args +\half\Delta\args +\half(\lambda^2-\phicl{2}{x})\delta_{m^2},\nn\\
\label{eqn:cps1}\\
c^0_{ps}\args&=\half(\partial_t^2+\partial_x\partial_{x'}+2)\,G^0\args +\half\Delta^0\args.  \label{eqn:c0ps}}
Note that we can set $t=0$  in \eqref{eqn:cps} because all the $\delta(t)$ components have cancelled once the subtraction on the r.h.s. has been computed. 
Using \eqref{eqn:cps}, \eqref{eqn:cps1} and \eqref{eqn:c0ps} 
gives
\AMSeqn{c_{ps}(x,x')&=\left.\half(\partial_t^2+\partial_x\partial_{x'}+\frac{3}{\lambda^2}\phicl{2}{x}-1)G^s\args +\half\Delta^s\args\right \vert_{t=0}\nn\\
&\quad+\threehalves\left(\frac{\phicl{2}{x}}{\lambda^2}-1\right)\left(G^0\argsO-\frac{\lambda^2}{3}\delta_{m^2}\right). \label{eqn:cdiff}}
The last term in \eqref{eqn:cdiff} vanishes by the definition of the mass counter-term \eqref{eqn:deltam}.  The point spectrum contribution  to $c_{ps}$ is
\AMSeqn{ c_{ps}^P(x,x')=&\left.\half\int \frac{d\omega}{2\pi} e^{i\omega(t-t')} \, \left(\frac{-\omega^2+\partial_x\partial_{x'}+ 3\lambda^{-2}\phicl{2}{x}-1}{\omega^2+\threehalves}+1\right) \psi_a(x)\psi_a(x') \right\vert_{t=0} \nn\cr
=& \quarter \sqrt{\frac{2}{3}}  
\left(\partial_x\partial_{x'}+ 3\lambda^{-2}\phicl{2}{x}+\half\right) \psi_a(x)\psi_a(x')  
 }
 which is finite when $x\to x'$.  
 The continuous spectrum contribution is
 \AMSeqn{ c_{ps}^C(x,x')=& \half\int \frac{d\omega dk}{(2\pi)^2}e^{i\omega(t-t')}      \nn\\ &\qquad \left.\left(\frac{-\omega^2+\partial_x\partial_{x'}+ 3\lambda^{-2}\phicl{2}{x}-1}{\omega^2+k^2+2}+1\right)\left(\psi_k(x)\psi^*_k(x')-e^{ik(x-x')} \right) \right\vert_{t=0}  \nn\cr
=& \quarter \int \frac{dk}{2\pi\sqrt{k^2+2}}    
\left(\partial_x\partial_{x'}+k^2+ 3\lambda^{-2}\phicl{2}{x}+1\right) 
\left(\psi_k(x)\psi^*_k(x')-e^{ik(x-x')} \right).\label{eqn:cC}
 }
Using the explicit form \eqref{eqn:psikdef} for $\psi_k(x)$  gives
\begin{equation}
    c_{ps}^C(x,x')=\sum_{l=0}^5B_l(x,x')\mathcal K_l(x,x'),
\end{equation}
where 
\begin{equation}
    \mathcal K_l(x,x')= \int\frac{k^l\,dk}{16\pi}\frac{e^{ik(x-x')}}{(k^2+2)^{3/2}(2k^2+1)}, \label{kell}
\end{equation}
and the functions $B_\ell(x,x')$ are regular at $x=x'$. As the $\cK_{\ell}(x,x')$ are finite  at $x=x'$ for $\ell<4$, it is sufficient to evaluate them there,
\begin{align}
    \mathcal K_0(x,x)&=\frac{4 \sqrt{3} \pi -9}{432 \pi },\nn\\
    \mathcal K_1(x,x)&=\mathcal K_3(x,x)=0,\nn\\
    \mathcal K_2(x,x)&=\frac{9-\sqrt{3} \pi }{216 \pi }.
\end{align}
For $l=4,5$, we have,\footnote{To show this, we expand $k^4(2k^2+1)^{-1}$, in \eqref{kell}, in powers of $(k^2+2)^{-1}$ which gives
\begin{equation}
    \mathcal K_4(x,x')=\frac{1}{32\pi}\int dk\,e^{ik(x-x')}\left\{\frac{1}{(k^2+2)^{3/2}}-{O}\left(\frac{1}{(k^2+2)^{5/2}}\right)\right\}.
\end{equation}
The sub-leading terms give a finite contribution when $x'\to x$, so we will not be interested in these; the leading term, on the other hand, is divergent when $x'\to x$ and is given by $\frac{1}{16\pi}K_0((x-x')/\sqrt 2)$ where $K_0$ is a modified Bessel function of the second kind. $\cK_5$ is then obtained by acting with $-i\partial_x$ on $\cK_4$.} 
\begin{align}
    \mathcal{K}_4(x,x')&=-\frac{\log|x-x'|}{16\pi}\left(1+{O}(x-x')\right),\\
    \mathcal{K}_5(x,x')&=\frac{i}{16\pi(x-x')}+{O}(1).
\end{align}
Although $\cK_4$ is logarithmically divergent, $B_4(x,x')={O}(x-x')$, so there is no $\ell=4$ contribution to $\expect{c(x)}$. However there is a contribution for $\ell=5$ because
\begin{align}
    B_5(x,x')&=12i\,\text{sech}^2\left(\frac{x}{\sqrt 2}\right)(x-x')+{O}\left((x-x')^2\right),
\end{align}
and the linear divergence in $\cK_5$ exposes the leading term. Finally, using
\begin{align}
     B_0(x,x')&=-3\text{sech}^2\left(\frac{x}{\sqrt 2}\right) \left(1+15\tanh^4\left(\frac{x}{\sqrt 2}\right)\right)+{O}(x-x'),\nn\\
    B_2(x,x')&=3\text{sech}^2\left(\frac{x}{\sqrt 2}\right)\left(1-9\tanh^2\left(\frac{x}{\sqrt 2}\right)\right)+{O}(x-x'),
\end{align}
we find 
\begin{align}
    \langle c(x)\rangle&=\lim_{x'\to x} \left(c_{ps}^P(x,x')+c_{ps}^C(x,x')\right),\nn\\
    &=c_{ps}^P(x,x)+\sum_{l=0,2}B_l(x,x)\mathcal{K}_l(x,x)+\lim_{x'\to x}B_5(x,x')\mathcal{K}_5(x,x')\nn\\
    &=\frac{\text{sech}^6(\frac{x}{\sqrt{2}})}{96 \pi }\left( (-72-9\sqrt{3} \pi ) \cosh (\sqrt{2} x)+(2 \sqrt{3} \pi -9) \cosh (2
   \sqrt{2} x)+ 27+9\sqrt{3} \pi \right). \label{naww1}
\end{align}
The moments are
\begin{equation}
    c_{(0)}=\frac{1}{2\sqrt 6}-\frac{3}{\pi\sqrt 2}, \quad\quad\quad
    c_{(2)}=\frac{\pi  \left(14 \sqrt{3}+\pi  \left(\sqrt{3} \pi -18\right)\right)-18}{36 \sqrt{2} \pi }.
\end{equation}
\noindent After mass renormalization, the final contribution to \eqref{eqn:gamma} contains only convergent $k$ integrals so point splitting is not necessary. We find 
\begin{align}
    &-2M_0^2\,\int \langle S_{int}\,\chi(x)\rangle\, \phi_{cl}'(x)\, x\,dx \nonumber\\
    &=6M_0^2\lambda^{-2} \int x\, dt\,dy\,dx\,\phi_{cl}'(x)\phi_{cl}(y)G^s(0;y,y)G(t;y,x)\nonumber\\
    &=\frac{M_0^2(-9-8 \sqrt{3} \pi +\sqrt{3} \pi ^3)}{18 \sqrt{2} \pi }.
\end{align}

\section{Discussion}

Combining the results obtained above, 
restoring factors of $m$, and introducing the dimensionless quantity $\bar\lambda\equiv(m\lambda)^{-1}$, 
the kink mass \eqref{eqn:mass} is given by  
\begin{align}
    Mm^{-1}=\frac{2\sqrt{2}}{3}\bar\lambda^{-2}\left(1+\bar\lambda^{2}\underbrace{\left(\frac{\sqrt 3}{8}-\frac{9}{4\pi}\right)}_{\approx -0.50}+{O}(\bar\lambda^{4})\right),\label{eqn:Massresult}
\end{align}
in agreement with the standard result \cite{DHN1974,Wiedig,Raj}. The radius $R$ \eqref{curlym} is given by
\begin{equation}
    mR=\sqrt{\frac{1}{6} \left(\pi ^2-6\right)}\left(1+\bar\lambda^{2} \underbrace{\frac{\sqrt{3} \pi  \left(2+\pi ^2\right)-72}{8 \pi  \left(\pi ^2-6\right)}}_{\approx -0.076}+{O}(\bar\lambda^{4})\right),\label{eqn:Radiusresult}
\end{equation}
or by 
\begin{equation}
    MR=\frac{2\bar\lambda^{-2}}{3}\sqrt{\frac{1}{3} \left(\pi ^2-6\right)}+\underbrace{\frac{ \left(\pi ^2-2\right)(\pi-3\sqrt 3)}{6 \pi  \sqrt{ \left(\pi ^2-6\right)}}}_{\approx -0.44}
   +{O}(\bar\lambda^{2}).
\end{equation}
In the perturbative, $0<\bar\lambda\ll 1$, regime,  $R$ is very much greater than the Compton wavelength of the soliton so it makes sense to regard the soliton as an extended object with the characteristics of a bound state. In this regime the topologically trivial excitations of the scalar field are scalar elementary (ie point-like) particles with a mass $m\ll M$; despite its excess mass the soliton in its ground state is stable against particle emission because it is topologically protected.
As $\bar\lambda$ increases, keeping $m$ fixed the scalar potential becomes more attractive so the binding energy increases which we expect to see manifested in reduced $M$, and size, $R$, of the soliton, and this is what the perturbation theory result shows. 

According to the Bound State Conjecture \cite{Hebecker2019}, if we can regard the soliton as a bound state, we expect $mR\geq\delta>0$ for all couplings. At least at first order in the perturbation expansion the relative size of the quantum correction for $M$ \eqref{eqn:Massresult} is seven times larger  than that for $R$ \eqref{eqn:Radiusresult}. Taking this literally, 
the solitons will become very light while $mR$ is still $O(1)$.  At this point we might expect the creation of kink / anti-kink pairs to become important and the state cease to be a single identifiable kink; as these pairs can separate without energy cost, the decrease in $R$ will be turned off in a manner that is consistent with the conjecture. 
Clearly, we would not be able to make this argument if the relative size of the quantum correction for $R$ were to be larger than for $M$. 

The expansion in $\bar\lambda$ gives us a clue about what might happen as the coupling strength increases. However to show that the mechanism described above actually prevents the soliton from violating the conjecture of \cite{Hebecker2019} we need to go beyond perturbation theory to include a full quantum description of kinks and anti-kinks; 
in which  multi-soliton matrix elements  and pair creation of solitons 
are calculationally accessible. The question is probably best pursued in an integrable theory such as the sine-Gordon model.

\acknowledgments

PDX acknowledges the support of an Oxford Berman Scholarship. 
This research was funded in whole, or in part, by Research England. For the purpose of Open Access, the authors have applied a CC  BY public copyright licence to any Author Accepted Manuscript version arising from this submission.


 \bibliographystyle{JHEP} 

\bibliography{refs2.bib} 

\end{document}